\documentclass[11pt]{amsart}
\usepackage{geometry}                % See geometry.pdf to learn the layout options. There are lots.
\usepackage{graphicx}
\usepackage{amssymb}
\usepackage{epstopdf}
\newcommand{\R}{\mathbb{R}}
\newcommand{\Z}{\mathbb{Z}}
\newcommand{\T}{\mathbb{T}}
\newtheorem{thm}{Theorem}
\newcommand{\eps}{\varepsilon}

\title{Surfaces of locally minimal flux}
\author{R.S.MacKay}
\address{Mathematics Institute, University of Warwick, Coventry CV4 7AL, U.K.}
\email{R.S.MacKay@warwick.ac.uk}
\date{\today}                                           % Activate to display a given date or no date

\begin{document}

\begin{abstract}
For exact area-preserving twist maps, curves were constructed through the gaps of cantori in \cite{MMP84}, which were conjectured to have minimal flux subject to passing through the points of the cantorus.  It was pointed out by \cite{Pol} that these curves do {\em not} have minimal flux if there coexists a rotational invariant circle of a different rotation number, but if hyperbolic they do have {\em locally} minimal flux even without the constraint of passing through the points of the cantorus.  Following the criterion of \cite{M94} for surfaces of locally minimal flux for 3D volume-preserving flows, I revisit this result and show that in general the analogous curves through the points of rotationally-ordered periodic orbits or their heteroclinic orbits do {\em not} have locally minimal flux.  Along the way, various questions are posed.  Some results for more degrees of freedom are summarised.
\vskip 1ex
In memory of John Mather (9 June 1942 -- 28 Jan 2017).
\end{abstract}

\subjclass[2010]{}

\maketitle

\section{Introduction}

An {\em exact area-preserving twist map} is a $C^1$ map $f:(x,y)\mapsto (x',y')$ of a cylinder $\T\times \R$ (where $\T\ = \R / \Z$) that preserves the standard area-form $dx\wedge dy$, such that for one or any lift $\tilde{f}$ to $\R\times\R$ the image of each vertical $x=$ constant is a $C^1$ graph over $x'$ (twist condition), and the nett flux $\int_\gamma y' dx' - y dx$ for any homotopically non-trivial circle $\gamma$ is zero.  We usually fix a lift and often drop the tilde.  

There exists a $C^2$ function $h:~\R\times\R \to \R$, called {\em generating function}, with $h_{12}<0$ and $h(x+1,x'+1)=h(x,x')$, such that for the lift $\tilde{f}$,
\begin{eqnarray}
y = -h_1(x,x') \\
y' = h_2(x,x'), \nonumber
\end{eqnarray}
where subscript $i$ denotes the derivative with respect to the $i^{th}$ argument.  It can be constructed by $h(x,x') = \int_\gamma y' dx' - y dx$ for any curve $\gamma$ from a base point $(x_0,x_0')$ to $(x,x')$, using the functions $y'(x,x')$ and $y(x,x')$ defined by the twist condition.
It follows that $(x_n,y_n)_{n\in\Z}$ is an orbit of $\tilde{f}$ iff for all $M<N \in \Z$, $(x_M,\ldots,x_N)$ is a critical point of $W_{MN}(x) = \sum_{n=M}^{N-1} h(x_n,x_{n+1})$ subject to $x_M,x_N$ fixed, and $y_n = -h_1(x_n,x_{n+1})$.  An orbit is said to be {\em minimising} if the sequence $(x_n)_{n\in\Z}$ (globally) minimises $W_{MN}(x)$ for each $M<N$ over variations with fixed endpoints.

For exact area-preserving twist maps $f$ of a cylinder $\T \times \R$ with generating function $h: \R\times \R \to \R$, Aubry-Mather theory (e.g.~\cite{KH}) establishes the existence for each rational $p/q$ (in lowest terms) of a closed rotationally-ordered set $M_{p/q}$ of periodic points $(x_n,y_n)$ of type $(p,q)$ (meaning $x_q = x_0 + p$, $y_q=y_0$ in the cover) that (globally) minimise the action $W_{p,q} = \sum_{n=0}^{q-1} h(x_n, x_{n+1})$.
In each gap in $M_{p/q}$, it also establishes the existence of a minimax orbit, which forms a saddle point of the action between the consecutive minima.  More precisely, denote the sequences of type $(p,q)$ for the endpoints of a gap $g$ in $M_{p/q}$ by $x^- \ll x^+$ (where $\ll$ means each component on the left is less than the corresponding one on the right), let $X_g$ be the set of sequences $x$ of type $(p,q)$ with $x^-_n\le x_n \le x^+_n$, and let $W_g$ be $W_{p,q}$ on $X_g$; let $H_g$ be the infimum of $H \in \R$ such that $x^-$ and $x^+$ lie in the same connected component of $\{x \in X_g : W_g(x) \le H\}$ (note that the endpoints have the same value of $W_g$).  Mather proved there is a critical point $x$ of $W_g$ in the interior of $X_g$ with $W_g (x)=H_g$.  The difference $\Delta W_g = H_g - W_g(x^-)$ is known as the Peierls-Nabarro barrier for shifting a type $(p,q)$ sequence between the consecutive pair of  minima while maintaining type $(p,q)$.

For any gap (in the rotational order) in the set of minimising periodic orbits of type $(p,q)$, Aubry-Mather theory gives also the existence of heteroclinic orbits from the orbit of the left end of the gap to that of the right end, and vice versa, that are minimising, and associated minimax heteroclinic orbits.  

Finally, for each irrational $\omega$, it gives a closed rotationally-ordered invariant set with rotation number $\omega$ whose orbits minimise the action sum between any pair of its points.  Its subset of recurrent points is either a circle or a Cantor set, the latter case being christened ``cantorus'' by Percival \cite{Per}.
The gaps of a cantorus come in orbits.  We call an orbit of gaps a {\em hole}.  The number of holes in a cantorus is countable (a question is whether it is generically finite).  For each hole $g$ in a cantorus, an analogue of $\Delta W_g$ is defined and an associated minimax orbit $M$ proved to exist \cite{Mat}, with
\begin{equation}
\Delta W(M,m) := \sum_{n\in \Z} h(M_n,M_{n+1}) - h(m_n,m_{n+1})
\label{eq:DeltaW}
\end{equation}
equal to $\Delta W_g$, where $m$ is the sequence for either of the minimising orbits bounding the hole.  The sum $\Delta W(M,m)$ converges even though the individual sums do not in general.  
The orbits of the endpoints of a hole converge together in both directions of time and the minimax orbit lies between them, so it is homoclinic to the cantorus. $\Delta W_g \ge 0$ with equality iff there is an arc of minimising orbits connecting the endpoints of the gap (in which case those minimising orbits are all minimax).

For any rotational (i.e.~homotopically nontrivial) closed curve $\gamma$ around the cylinder, the nett flux $\int_{f(\gamma)} y\ dx - \int_\gamma y\ dx$ is zero, but we define its {\em flux} to be the area of the set lying in the component below $f(\gamma)$ and not in the component below $\gamma$, sometimes called {\em geometric flux}.  

\begin{figure}[htbp] %  figure placement: here, top, bottom, or page
   \centering
   \includegraphics[width=5.5in]{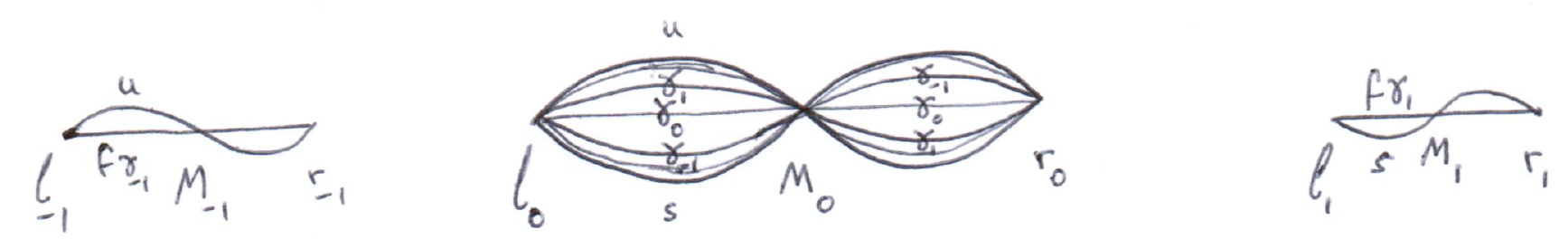} 
   \caption{\small Decomposition of a turnstile.}
   \label{fig:turnstile}
\end{figure}

For a rotational invariant circle, the flux is zero.  For a cantorus of rotation number $\omega$, one can close its gaps by curves through corresponding minimax orbits, to produce what \cite{MMP84} called a {\em partial barrier}.  For example, if the cantorus is hyperbolic, for each hole one can choose a $0^{th}$ gap and close it and its forward images by arcs of stable manifold of the cantorus and its backward images by arcs of unstable manifold.\footnote{The stable and unstable manifolds are better called forward-contracting and backward-contracting manifolds but I stick with convention for shorter names.}
These automatically join the ends of the gaps and pass through the minimax points.  This was the choice made in \cite{MMP84}, though it was indicated there that there is a wealth of other options.  In particular, \cite{MMP84} defined a {\em turnstile} in the $0^{th}$ gap as the structure formed by the arc of stable manifold of the partial barrier and the arc of unstable manifold whose backward images were used to close the backward gaps.
In the simplest case when the stable and unstable manifolds in each gap intersect precisely once, thus in the minimax point, the turnstile can be sliced into slivers by disjoint curves $\gamma_n, n\in \Z$, joining the ends (like the layers of an onion), and the $\gamma_n$ can be mapped by $f^n$ to the $n^{th}$ gap to make a curve closing the gaps and having the same flux (see Figure~\ref{fig:turnstile}).
%given one curve through the points of a cantorus and minimax orbits, one can flow it along any vector field that is zero on the cantorus and minimax orbits to get another curve, and doing this judiciously one preserves the flux.
 
In this simplest case, the flux of such a curve, whether the original partial barrier or one as constructed by the onion picture, is $$\widetilde{\Delta W}_\omega = \sum_g \Delta W_g,$$ where the sum ranges over the holes $g$ of the cantorus \cite{MMP84}.
Mather defined $\Delta W_\omega$ for a cantorus to be the maximum of $\Delta W_g$ over its holes $g$ \cite{Mat}.   We see here that the {\em sum} over holes has a valuable interpretation, because it gives the flux for a closed curve through all the points of the cantorus (note that the sum over holes converges even if there are infinitely many, because the gaps are disjoint).  Mather was surprisingly (to me) uninterested in replacing the maximum by the sum when I proposed this to him in 1984.  I suspect that the continuity properties he proved for the maximum apply equally well to the sum, but it would be good to check this.
He was nevertheless interested in the question of motion within a Birkhoff zone of instability, notably proving existence of an orbit whose $\alpha$-limit set is in the lower boundary and $\omega$-limit set in the upper boundary \cite{Mat2}.

\begin{figure}[t] %  figure placement: here, top, bottom, or page
   \centering
   \includegraphics[width=4in]{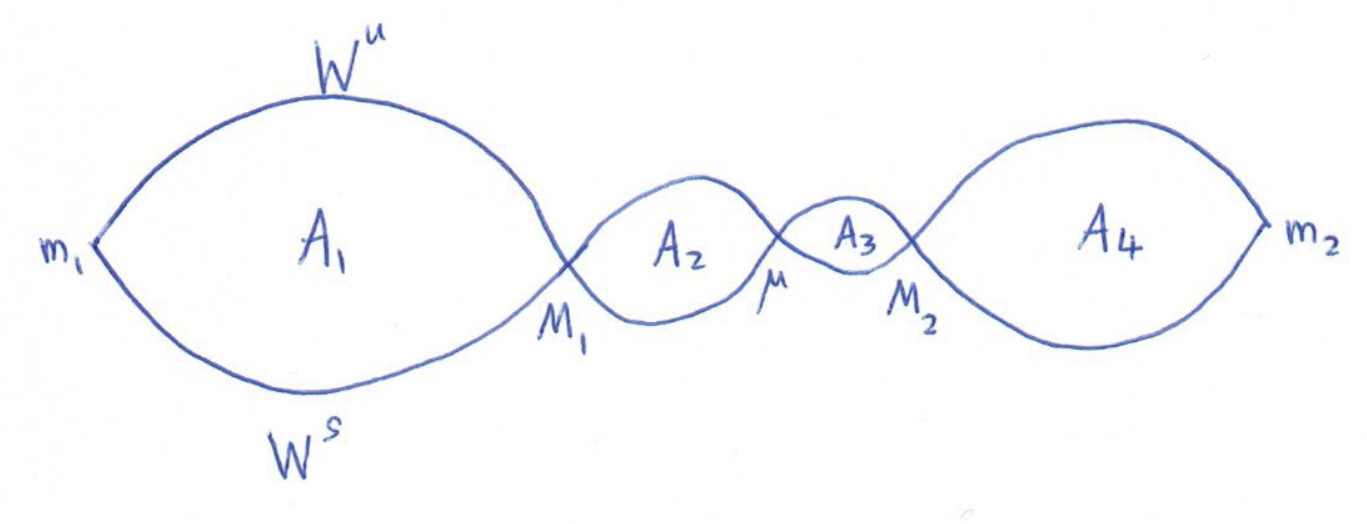} 
   \caption{\small An example of a gap in a cantorus for which the arcs of stable and unstable manifold closing the gap intersect more than once.  The endpoints of the gap are $m_1, m_2$.  The areas $A_i$ satisfy $A_2 < A_1$, $A_3 < A_4$, and $A_1+A_3 = A_2 + A_4$, the common value being the flux of the set of curves formed by iterating the unstable manifold backwards and the stable manifold forwards.  If $A_1>A_4$, as shown here, then the minimax point is $M_1$, so Mather's $\Delta W = A_1$.  The point $M_2$ is a lower saddle, but to get from the orbit of $m_1$ to the orbit of $m_2$ one has to pass at least as high as $M_1$ in action.  The point $\mu$ belongs to a local minimum orbit with higher action than $m_1$ and $m_2$ (which have the same action).}
   \label{fig:multipleintersections}
\end{figure}
\begin{figure}[b] %  figure placement: here, top, bottom, or page
   \centering
   \includegraphics[width=2.7in]{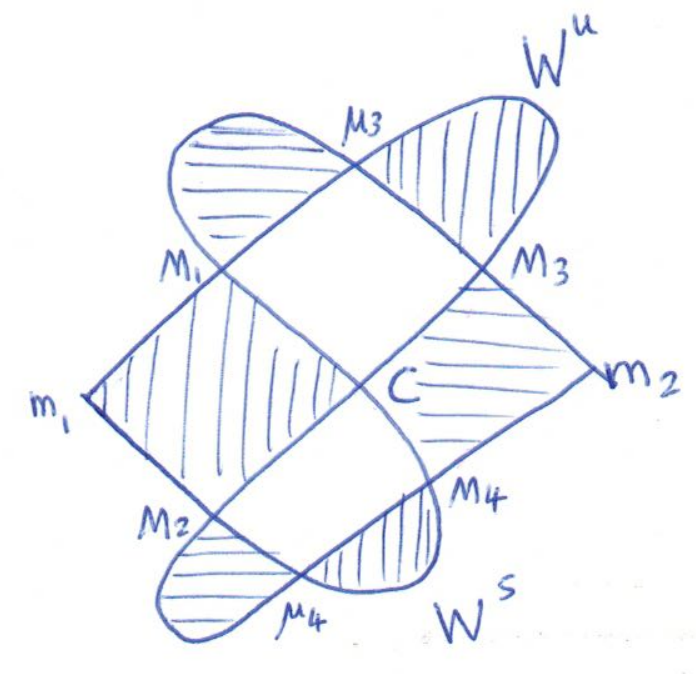} 
   \caption{\small A more complicated scenario for the stable and unstable manifolds of the endpoints $m_1, m_2$ of a gap in a cantorus.  The upward flux of the set of curves formed by iterating the unstable manifold backwards and the stable manifold forwards is the vertically shaded area and the (equal) downward flux is the horizontally shaded area.  One of the points $M_i$ is a minimax point, depending which is the lowest saddle permitting to go from $m_1$ to $m_2$ without passing to higher action.  The points $\mu_i$ belong to local minima of the action.  The point $C$ belongs to an index-2 critical point of the action.}
   \label{fig:notgraphs}
\end{figure}

More complicated scenarios can occur, however, which might explain Mather's disinterest.  One way to construct such examples is to use Aubry's idea of the anti-integrable limit \cite{BM}.
Firstly, there may be other equilibrium sequences in a hole besides the chosen minimax, as in Figure~\ref{fig:multipleintersections}.  The stable and unstable manifolds intersect at each of them.  
If the stable and unstable manifolds are graphs in the gap, or even just if there is a homeomorphism that makes them both graphs, then the flux is the sum of differences of action between alternating successive pairs of equilibrium sequence, which is larger than $\widetilde{\Delta W}_\omega$.  
A good question, however, is whether there is an alternative way of closing the gaps for which the flux is just $\widetilde{\Delta W}_\omega$.

Secondly, one can make cases where the stable and unstable manifolds in a gap are not simultaneously homeomorphic to graphs, like that of Figure~\ref{fig:notgraphs}.  In the case shown, there are seven equilibrium sequences in the hole (plus the two endpoints) and the flux is given by 
\begin{equation}
\Delta W(M_2,m) - \Delta W(C,M_1) + \Delta W(M_3,\mu_3) + \Delta W(M_4,\mu_4),
\end{equation}
where the notation $\Delta W$ is extended to arbitrary pairs of equilibrium sequences converging together sufficiently fast to make the sum (\ref{eq:DeltaW}) converge.
 
%In the exceptional case that the set of minimisers forms a circle then the arcs of the circle close the gaps in an invariant way, producing flux zero.  

\cite{MMP84} conjectured that the curves $\gamma$ closing the gaps of hyperbolic cantori by arcs of stable or unstable manifold have minimal flux subject to passing through the cantorus (the local minimality or otherwise of the flux of such curves was also addressed by \cite{BK}).  However, \cite{Pol} proved that this is false if there coexists a rotational invariant circle (necessarily of a different rotation number), because large deformations of such curves can be made to obtain curves through the cantorus with arbitrarily small flux.  Nevertheless, he proved that the curves $\gamma$ have locally minimal flux, even without the constraint of passing through the points of the cantorus.  This makes them valuable for transport theory.

Nevertheless, there are alternative constructions of curves closing the gaps of cantori, and the question arises whether they also have locally minimal flux.  We address that here.  Secondly, there are related constructions of curves through minimising and minimax periodic orbits and through minimising and minimax heteroclinic orbits to minimising periodic orbits \cite{MMP84, BK, MMP2}, and the question arises whether they have locally minimal flux too.  The answer is no:~they do not have locally minimal flux, a result that I am not aware has been noted before.  Thirdly, the issues arise equally in continuous-time Hamiltonian systems of $1\frac12$ degrees of freedom (DoF), and more generally in 3D volume-preserving flows.  We address these contexts.  Fourthly, we summarise some related results for Hamiltonian systems of more DoF.

\section{Other constructions of partial barriers}

Various other constructions of curves closing the gaps of cantori can be used.  I survey two here.  One is Hall's ``ridge curves", the other is Dewar's ``quadratic flux minimisers''.

\subsection{Ridge curves}
Ridge curves were proposed by Hall (private communication) and popularised under the name ``ghost circles" by Gol\'e \cite{Gol}.  A nice description is via the ``invariant ordered circles" (IOC) of \cite{QW}.  An {\em ordered circle} is a continuous curve $x: \R \to \R^\Z$ that is ordered (for each pair of distinct $t, t' \in \R$ either $x_n(t)<x_n(t')$ for all $n\in \Z$ or vice versa), periodic (invariant under the translation $T_{01}(x)_n = x_n + 1$), and unbounded (there is no $y \in \R^\Z$ such that $x(t) \le y$ for all $t\in\R$, nor $x$ such that $x \le x(t)$ for all $t\in \R$).  It is called {\em invariant} if it is also invariant under the translation $T_{10}(x)_n = x_{n+1}$ and the gradient flow of the action:
\begin{equation}
\dot{x}_n = -h_2(x_{n-1},x_n) - h_1(x_n,x_{n+1}).
\end{equation}
Note that the gradient flow, $T_{01}$ and $T_{10}$ commute.

\cite{Gol} constructs IOCs firstly in the space of sequences of type $(p,q)$ by assuming the action $W_{p,q}$ is a Morse function and using minimax theory iteratively, and then takes limits for the irrational and heteroclinic cases.  \cite{QW} construct them by the Schauder fixed point theorem for the time-1 map of the gradient flow, which is more elegant as it does not require the Morse assumption and applies equally well to periodic and irrational cases.  

However constructed, they produce curves that close the gaps of cantori and a nice flux formula \cite{Gol}, as follows.
If we let 
\begin{eqnarray}
y^+_n &=& h_2(x_{n-1},x_n) \\
y^-_n &=& - h_1(x_n,x_{n+1}),
\end{eqnarray}
then each IOC produces two rotational circles around the cylinder: $\gamma_n^- = (x_n(t),y^-_n(t))$ and $\gamma_n^+(x_n(t), y^+_n(t))$, where $t$ is a parameter along the IOC.  Both are graphs over $x$.  Furthermore, $f(\gamma_n^-) = \gamma_{n+1}^+$.  We see that $\gamma_n^+$ is above $\gamma_n^-$ at places where $\dot{x}_n < 0$ and below at places where $\dot{x}_n > 0$.
The dynamics of the gradient flow (and its time-reverse) are monotone, so if $x$ is an initial condition for which $\dot{x}_n <0$ for all $n\in \Z$ (we write $\dot{x} \ll 0$) then it remains so for all $t\in \R$.  Ridge curves are curves in the space of sequences consisting of gradient curves on which $\dot{x} \gg 0$ or $\dot{x} \ll 0$, joined at equilibria.  By the above construction, they produce pairs $(\gamma^-,\gamma^+)$ of graphs on the cylinder with $f(\gamma^-)=\gamma^+$ such that for each interval where $\dot{x}\ll 0$ then $\gamma^+$ is above $\gamma^-$ and for each interval where $\dot{x}\gg 0$ then $\gamma^+$ is below $\gamma^-$.  They intersect at orbits corresponding to the equilibrium sequences.  Thus the flux of $\gamma^-$ is produced precisely by its intervals with $\dot{x}\ll 0$.  Furthermore the contribution of such an interval to the flux is precisely the difference in action $\Delta W(r,l)$ between the orbits of the equilibrium sequences $r,l$ at its right and left ends.
Thus the flux of $\gamma^-$ is the sum of $\Delta W (r,l)$ over those intervals $(l,r)$ with $\dot{x}\ll 0$, which is Gol\'e's formula.  

The construction can be generalised.  It is not necessary for the ordered circle to be invariant under the gradient flow.  All that we need is for $\dot{x}\ge 0$ or $\dot{x}\le 0$ in the standard partial order on $\R^\Z$ at each point.

%??? Another construction, motivated directly by the aim of producing curves of locally minimal flux, is quadratic flux minimisation \cite{DM}, perhaps better called $L^2$-flux minimisation.  It applies to continuous-time Hamiltonian systems rather than maps.  
%Consider the case of $H(x,y,t)$, periodic in $x$ and $t$, with $\dot{x} = H_y$, $\dot{y} = -H_x$ (subscripts representing partial derivatives).  For a graph $y=Y(x,t)$ define the quadratic flux to be

\subsection{Quadratic flux minimisers}
Dewar didn't like the non-uniqueness of local flux minimisers and proposed an $L^2$ variational principle instead, to select a preferred $L^1$ minimiser.  He first did this to construct a preferred action variable for a non-integrable Hamiltonian system.  In the context of area-preserving maps, it was worked out with Meiss in \cite{DM}.  The idea is that the flux of an area-preserving map $f$ across a closed curve $\gamma$ around the cylinder can be written as $\frac12 \int |\Delta y(x)|\ dx$ if $\gamma$ and $f(\gamma)$ are both graphs of functions $y_0, y_1$ of $x$ and $\Delta y(x) = y_1(x)-y_0(x)$.  Minimising this is an $L^1$ variational problem.   To overcome the non-uniqueness, they proposed instead to minimise something like $\int |\Delta y(x)|^2\ dx$, with the hope that there is a local minimiser associated to each rotation number and that the resulting curves $\gamma$ are disjoint for different rotation numbers.

The approach is somewhat analogous to that for geodesics in a Riemannian manifold.  They are defined as curves for which every short enough segment minimises the $L^1$ functional $\int_{t_0}^{t_1} |\dot{x}(t)|\ dt$ subject to fixed endpoints $x(t_0) = x_0, x(t_1)=x_1$.  But the minimisers can be reparametrised by any increasing function of $t$ fixing the ends, which does not change the value of the integral.  The $L^2$ variational principle of minimising $\int |\dot{x}|^2\ dt$ selects a preferred parametrisation, namely such that $|\dot{x}(t)|$ is constant.  The relation between the two variational principles is nicely summarised in section 12 of \cite{Mi}.

The variational principle in \cite{DM} is slightly more subtle than sketched above.  In fact they proposed to locally minimise $\int |\Delta y(x)|^2\ x'(\theta)\ d\theta$ over pairs $(x,\rho)$ of increasing diagonally periodic homeomorphisms (i.e.~$x(\theta+1)=x(\theta)+1$ and the same for $\rho$), with the functions $y_0, y_1$ determined by
\begin{eqnarray}
y_0(x(\theta)) &=& -h_1(x(\theta),x(\rho(\theta))) \\
y_1(x(\theta)) &=& h_2(x(\rho^{-1}(\theta),x(\theta)),
\end{eqnarray}
so
\begin{equation}
\Delta y(x(\theta)) = h_2(x(\rho^{-1}(\theta),x(\theta))+h_1(x(\theta),x(\rho(\theta))) .
\end{equation}
The nice result is that the Euler-Lagrange equations for stationarity of the $L^2$ functional implies that if $\Delta y(x(\theta))=0$ for some $\theta$ then $\Delta y(x(\rho^n(\theta)))=0$ for all $n\in \Z$.  Thus the $L^2$ critical points make a curve $\gamma$ and its image $f(\gamma)$, whose intersections are orbits of $f$.  So the $L^2$ variational principle constructs preferred curves through selected orbits of $f$.

Yet many questions remain (at least for me).  Does the principle have local minimisers?  Is there one for each rotation number?  Are the curves $\gamma$ for different rotation number disjoint?  Is there a relation between the $L^2$ minimisers and the ridge curves?  The papers \cite{DHG, DHG2, DHG3} address these questions but I'm not yet clear if they resolve them totally.

An alternative selection procedure among curves of locally minimal flux has been proposed by \cite{F}, based on minimising their length.

\section{Continuous-time analogues}
For many applications, rather than area-preserving maps it is better to consider continuous-time systems, e.g.~time-periodic Hamiltonian systems of $1\frac12$ DoF.
\begin{eqnarray}
\dot{x} &=& H_{,p}(x,p,t) \\
\dot{p} &=& -H_{,x}(x,p,t) \nonumber
\end{eqnarray}
with $H(x,p,t+1) = H(x,p,t)$ (where $H_{,p}$ denotes the partial derivative of $H$ with respect to $p$ etc.).  They preserve the volume-form $dx\wedge dp \wedge dt$ on the extended state space of $(x,p,t)$.  More generally one could consider 3D volume-preserving vector fields, such as a magnetic field $B$.

Let us take the time-periodic Hamiltonian context, with $x$ an angle variable.  The flux-form of the Hamiltonian vector field in extended state space $(x,p,t)$ is
$dH \wedge dt + dx \wedge dp$.
So the flux across a piece of a $C^1$ graph $p=P(x,t)$ is 
\begin{equation}
\int (H_{,x}+H_{,p}P_{,x}+P_{,t})\ dx \wedge dt.
\end{equation}
The nett flux across the whole graph is zero, but we define the flux of the surface to be the positive part, so the flux is $\frac12 \int |H_{,x}+H_{,p}P_{,x}+P_{,t}|\ dx\ dt$.

Now we can apply a result of \cite{M94}, which was derived in the more general context of 3D volume-preserving flows.
\begin{thm}
A surface has locally minimal flux for a 3D volume-preserving vector field iff it can be decomposed into surfaces of unidirectional flux bounded by trajectories and it has no local recrossings.
\end{thm}
\noindent Here, we say an oriented surface $S$ has {\em local recrossings} if for all $\eps > 0$ there exists an orbit segment $z(t), t_0 \le t \le t_1$, that intersects $S$ in opposite directions at times $t_0$ and $t_1$, and for which
$0 < d(z(t),S) < \eps$ for all $t \in [t_0,t_1]$, where $d$ denotes distance.  There is no requirement for $t_0$ and $t_1$ to be close.  The idea is close in spirit to Conley's concept of isolating block.

In particular, Aubry-Mather theory extends to time-periodic Hamiltonian systems with $x$ an angle and $H_{,pp} \ge C > 0$.  Cantori are now invariant subsets of graphs $p = P(x,t)$ where one or more infinitely long disjoint irrationally winding strips have been removed.  They consist of minimisers for the action functional $\int L(x,\dot{x},t)\ dt$ with $L(x,v,t) = \min_p (pv-H(x,p,t))$.
In each strip there is at least one minimax orbit.  If the cantorus is hyperbolic the stable and the unstable manifolds of the boundaries of each strip connect the boundaries and the minimax orbits.   A surface can be chosen between the invariant manifolds to fill in each strip, passing through the intersection orbits, and having unidirectional flux in between them.  One way to do this is to choose the unstable manifold up to $t=0$ and then switch with a vertical surface to the stable manifold for $t>0$, but this can be smoothed out if desired.
Along the lines of \cite{Pol}, it looks likely that the resulting surface has no local recrossings.  It would be good to write out a complete proof.  If so, then by the above theorem, hyperbolic cantori can be spanned by surfaces of locally minimal flux.

Note the corollary that every irrational $\omega$ is a local minimiser of the function $\widetilde{\Delta W}_\omega$.  This is because the cantorus of rotation number $\omega$ is a limit point of cantori of rotation numbers converging to $\omega$ \cite{Mat}.

\section{Surfaces through minimising and minimax periodic orbits}

We continue the discussion in the time-periodic Hamiltonian context of $1\frac12$ DoF with $H_{,pp}$ positive.  The set of minimising periodic orbits of type $(p,q)$ is rotationally ordered and closed.  For each gap in it there is a minimax periodic orbit.  Choose any closed surface that passes through the minimising and minimax orbits in rotational order.  Suppose we can choose it so that the parts between neighbouring minimax and minimising orbits have unidirectional flux.  Can one choose it so that it has no local recrossings?  Not in general, because the vector field rotates around any non-degenerate minimax periodic orbit, so there are arcs of trajectory arbitrarily close that cross the surface in one direction and recross in the other.
The flux can be reduced by pushing the surface off the minimax periodic orbit, analogous to \cite{BK}.

So the conclusion is that (except perhaps for degenerate cases) there are no surfaces of locally minimal flux through minimising and minimax periodic orbits.

\section{Surfaces through heteroclinic orbits to minimising periodic orbits}

In each gap of the set of minimising periodic orbits of type $(p,q)$ there is a set of minimising advancing heteroclinic orbits and in each gap of the latter there is a minimax heteroclinic orbit.  Advancing heteroclinic means that the orbit converges in forwards time to one periodic orbit and in backwards time to another and the forward limit is to the right of the backwards one. 
The same result holds for retreating heteroclinic orbits (for which the forward limit is to the left of the backward one), but without loss of generality, we will restrict the discussion to the advancing case. 

One can construct a surface through the minimising periodic orbits and their minimising advancing heteroclinic orbits, for example by taking the unstable manifold of a minimising periodic orbit up to some minimising heteroclinic orbit and the stable manifold on the other side, for one period of the periodic orbits and then closing by the required part of $t=cst$.  The only places where there is flux are the $t=cst$ pieces and it is unidirectional for the pieces between neighbouring pairs of heteroclinic orbits.  

\begin{figure}[htbp] %  figure placement: here, top, bottom, or page
   \centering
   \includegraphics[width=5in]{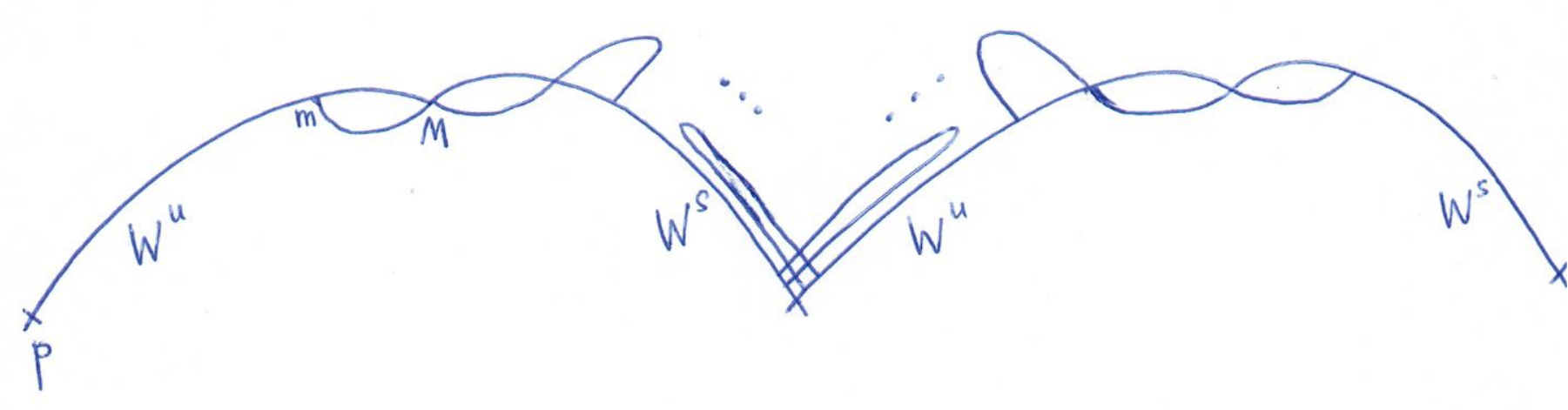} 
   \caption{\small Local recrossings happen for surfaces through hyperbolic minimising advancing heteroclinic orbits formed from stable and unstable manifolds.  $p$ is a minimising periodic point, $m$ and minimising advancing heteroclinic point and $M$ a minimax advancing heteroclinic point.  The picture is drawn at a Poincar\'e section $t=cst$.}
   \label{fig:minheteros}
\end{figure}

But local recrossings occur:~see Figure~\ref{fig:minheteros}.
So such surfaces do not have locally minimal flux.  This is consistent with 
%Mather's proof that there are cantori arbitrarily close and 
numerics (e.g.~Fig.11 of \cite{M83}) showing that there are cantori of smaller flux with arbitrarily close rotation number and lying in an arbitrarily small neighbourhood of the minimising periodic orbit.  But I am not aware of its having been remarked before.

\section{More degrees of freedom}

Finally, we summarise some results for Hamiltonian systems of more DoF.
An autonomous Hamiltonian system of $N$ DoF is specified by a $2N$-dimensional manifold $M$, a symplectic form $\omega$ on $M$ and a smooth function $H: M \to \R$.  The Hamiltonian vector field $v$ is determined by $i_v\omega=dH$.

Let the geometric flux for a codimension-1 closed surface in $H^{-1}(E)$ be the integral of the flux $i_v \sigma$ of energy-surface volume $\sigma$ (defined so that $dH \wedge \sigma = \omega^N/N!$) across the part where the flux is positive (the net flux = 0).  Note that
\begin{equation}
i_v \sigma = \omega^{(N-1)}/(N-1)!
\end{equation}

A {\em transition state} (TS) for an autonomous Hamiltonian system is a closed invariant oriented codimen\-sion-2 submanifold (not necessarily connected) of an energy level $H^{-1}(E)$ that can be spanned by compact codimension-1 surfaces of unidirectional flux whose union (dividing surface DS) locally separates $H^{-1}(E)$ into two components and has no local recrossings (recall this means there is a neighbourhood of DS that a trajectory has to leave before it can recross) \cite{MS}.

\begin{thm} (Theorem 2.3 of \cite{MS})
A codimension-1 closed submanifold in $H^{-1}(E)$ has locally minimal geometric flux iff it is a DS for a TS.
If $\omega = -d\alpha$ locally then the minimising flux is the action integral of the TS:~$ \int \alpha \wedge \omega^{(N-2)}/(N-2)!$.
\end{thm}
\noindent The result is formulated differently in \cite{MS} to motivate the definition of TS later, but it is equivalent to Theorem 2.3 there.

Examples can be constructed, in particular for energies just above an index-1 saddle of the Hamiltonian (these go back a long time in history, but see \cite{MS} for a coherent presentation).

A related variational principle for odd-dimensional invariant submanifolds of an autonomous Hamiltonian system, including the case of codimension-2 submanifolds of an energy level, was given in \cite{M91}, but with no general construction of minimisers.

\section{Potential application areas}
The theory of surfaces of locally minimal flux has many potential applications.  I mention two, to give an idea of the scope.  One is chemical reaction dynamics, as discussed in \cite{MS}.  Another is transport in magnetically confined plasmas, e.g.~\cite{W, C+, T+}, and the design of divertors.  To these one could add interplanetary travel and high energy particle storage rings.

%\section{Discussion}

\section*{Acknowledgements}
This work builds on results of John Mather that I learnt from him while a graduate student in Princeton.  It was a privilege to have him as a teacher.  It
was supported by the National Science Foundation under Grant No.~DMS-1440140 while the author was in residence at the Mathematical Sciences Research Institute in Berkeley, California, during the Fall 2018 semester.  I was prompted to write it up by a seminar of Bob Dewar there on surfaces of locally minimal flux.

\end{document}